\documentclass[10pt]{article} 
\makeatletter
\def\draft{\textheight=10.5truein \textwidth=7.5truein \parindent=8pt
           \voffset=-1truein \topmargin=0Truein
           \ifcase \@ptsize \hoffset=-1.5truein \or \hoffset=-1.35truein
                        \or \hoffset=-1.15truein \fi}
\def\quality{\textheight=240mm \textwidth=150mm \topmargin=0Truein
             \ifcase \@ptsize \hoffset=-23mm
                     \or \hoffset=-20mm \or \hoffset=-15mm \fi}
\makeatother \quality

\def\cB{{\cal B}} \def\cM{{\cal M}}  \def\SBR{{\bf SBR }}  \def\BV{{\bf BV}}
\def\cP{{\cal P}}   \def\Tor{{\bf Tor}}
\def\IR{\hbox{\rm I\kern-.2em\hbox{\rm R}}}  
\def\IC{{\bf C}}  \def\IL{{\bf L}}   \def\IP{{\bf P}}  \def\var{\, {\rm var}}
\def\IZ{\hbox{{\rm Z}\kern-.3em{\rm Z}}}
\def\const{\, {\rm Const} \,}  \def\mod1{\,({\rm mod\ } 1)\,}
\def\ep{\varepsilon}  \def\phi{\varphi}

\def\rmap{{\overline T}}
\def\cM{{\cal M}}  \def\La{\Lambda}  \def\la{\lambda}
\def\E#1{{\bf E}\left[#1\right]}
\def\lr#1{\left\{ #1 \right\}}
\def\CR{$$ $$}    \def\n{\noindent}

\def\nB#1{\norm{#1}{\cB}}  \def\sp{{\rm sp}}   \def\esssp{{\rm _{ess}sp}}
 
    \def\cF{{\cal F}}
\def\map{T}

\def\mlbscale{1pt} 
\def\bline(#1,#2)(#3,#4)(#5){\put(#1,#2){\line(#3,#4){#5}}}

\def\Bfig(#1,#2)#3#4{\begin{figure} \begin{center}
    \setlength{\unitlength}{\mlbscale} \begin{picture}(#1,#2)#3
    \end{picture} \end{center} \caption{ }#4 \end{figure}}

\def\bpic(#1,#2)#3{\setlength{\unitlength}{\mlbscale}
    \begin{picture}(#1,#2) #3 \end{picture}}

\newcount\myDx \newcount\myDy \newcount\myxy
\def\bdline[#1,#2](#3,#4)(#5,#6)(#7){  
    \myDx=#5 \multiply\myDx by #2   \myDy=#6 \multiply\myDy by #2
    \myxy=#7 \divide\myxy by #2  
    \multiput(#3,#4)(\myDx,\myDy){\myxy}{\line(#5,#6){#1}}
      \myDx=#5 \multiply\myDx by #7 \advance\myDx by #3
      \myDy=#6 \multiply\myDy by #7 \advance\myDy by #4
    \put(\myDx,\myDy){\line(-#5,-#6){#1}} }

\def\beq#1#2{\begin{equation} \label{#1} #2 \end{equation}}
\def\bea#1{\begin{eqnarray*} #1 \end{eqnarray*}} \def\a{\!\!\!&\!\!\!\!&}
\def\function#1{\left\{\!\!\!\begin{array}{ll} #1 \end{array} \right.}
\def\norm#1#2{\left|\left|#1\right|\right|_{#2}}      
\def\na#1#2{\norm{#1}{_{(#2)}}}
\def\n{\noindent}
\def\1#1{\hbox{\large \bf 1}_{#1}}     

\newtheorem{theorem}{Theorem}[section]   
\newtheorem{lemma}{Lemma}[section]       
\newtheorem{corollary}[lemma]{Corollary}      
\newtheorem{example}[lemma]{\exname}  \def\exname{Example}

\newtheorem{dftn}{Definition}[section]
\newenvironment{definition}{\begin{dftn}\rm}{\end{dftn}} 
\newtheorem{rmrk}[lemma]{Remark}

             \catcode`@=11 \@addtoreset{equation}{section} \catcode`@=12

\def\proof{\smallskip \noindent {\bf Proof. \ }}
\newcommand\filledsquare{\ \vrule width 1.5ex height 1.2ex}  
\def\qed{\hfill\filledsquare\linebreak\smallskip\par}

\begin{document}

\title{Perron-Frobenius spectrum for random maps \\
       and its approximation}
\author{Michael Blank\thanks{Russian Academy of Sciences, 
                             Inst. for Information Transmission Problems,
                             B.~Karetnij 19, 101447, Moscow, Russia,
                             and Observatoire de la Cote d'Azur, 
                             BP 4229, F-06304 Nice Cedex 4, France;
                             e-mail: blank@obs-nice.fr}
       }
\date{April 20, 2001}
\maketitle

\n {\bf Abstract.} To study the convergence to equilibrium in random 
maps we developed the spectral theory of the corresponding transfer 
(Perron-Frobenius) operators acting in a certain Banach space of 
generalized functions. The random maps under study in a sense fill 
the gap between expanding and hyperbolic systems since among their 
(deterministic) components there are both expanding and contracting 
ones. We prove stochastic stability of the Perron-Frobenius spectrum 
and developed its finite rank operator approximations by means of a 
``stochastically smoothed'' Ulam approximation scheme. A counterexample   
to the original Ulam conjecture about the approximation of the SBR 
measure and the discussion of the instability of spectral approximations 
by means of the original Ulam scheme are presented as well.

\section{Introduction}\label{sec:intro}

Let $\{\map_{i}\}$ be a collection of nonsingular maps from a
$d$-dimensional smooth manifold $X$ into itself with a metrics
$\rho$ on it, and let $\{p_{i}\}$ be a collection of nonnegative
integers, such that $\sum_{i}p_{i}=1$. A {\em random dynamical system} 
on the space $X$ is a stationary stochastic process 
$\map_1,\map_2,\cdots: X\to X$ (i.e., with values in a space of maps), 
see for instance \cite{Kif}. There are several approaches in further 
detalization of this object and in the sequel we shall use the following 
Markov one. 

By a {\em random map} $\rmap$ we shall mean the Markov random process 
on $X$ given by the following family of transition probabilities 
$\cP(x,A) := \sum_{i} p_{i} \1{\map_{i}^{-1}(A)}(x)$
from a point $x\in X$ to a subset $A\subseteq X$.
In other words, on every time step the map $\map_{i}$ is chosen
from the collection $\{\map_{i}\}$ independently from the
previous choices with the probability given by the distribution
$\{p_{i}\}$.

In the literature (especially physical) the above defined random
map is often called by the {\em iterated function system} (with
probabilities). Naturally, a pure deterministic setup was studied
as well. This can be done as follows. According to the given
collection of maps $\{\map_i\}$ one can define a new multivalued
map $\tilde\map:X\to X$ as $\tilde\map x:=\cup_i\map_ix$.
Assuming that the maps $\map_i$ are continuous and strictly
contracting one can show \cite{Hut} that the multivalued map
$\tilde\map$ possesses a global attractor $X_{\tilde\map}\subset
X$, namely the Hausdorff distance between the sets
$\tilde\map^nY$ and $X_{\tilde\map}$ decreases exponentially fast
for any closed nonempty set $Y\subset X$.

It is worth note that our aim in this paper is not to study the
most general setup (for example, one can consider an infinite
(continual) collection of maps $\{\map_{i}\}$, while the
distribution $\{p_{i}\}$ might be place dependent (on the space variable), 
but rather to give a deeper analysis of dynamical and statistical 
properties of these systems related to the convergence to equilibrium, 
i.e. on spectral problems of the corresponding transfer operators
practically not studied in the literature (except \cite{Bl-mon}).
Therefore we shall not consider well studied problems of the dimensional 
analysis of invariant sets and measures. The reader can find results and 
further references of this type, e.g. in \cite{Fal}.

One can always realize the random map under study as a
deterministic one on the extended phase space. Denote by
$\bar\Omega$ the space of one-side sequences
$\bar\omega:=\{\omega_{1},\omega_{2},\dots\}\in\bar\Omega$, where
each $\omega_{i}$ belongs to the set of indices of the collection
$\map_{i}$. On the space $\bar\Omega$ one defines the left-shift
map $\sigma:\bar\Omega\to\bar\Omega$ according to the rule
$(\sigma\bar\omega)_{i} := \omega_{i+1}$. The topology in the
space $\bar\Omega$ is defined as the direct product of discrete
topologies, acting on the set of indices, and a Borel measure
$\mu_{p}$ on $\bar\Omega$ is defined as the product of
distributions $\{p_{i}\}$ on the set of indices. Naturally there
exists an one-to-one correspondence between the sequences of maps
$\map_{i}$ and the elements of the space $\bar\Omega$. On the
extended space $\bar\Omega\times X$ one can define a new map -- 
a skew product upon the left-shift map: $\tilde\map(\bar\omega,x) :=
(\sigma\bar\omega, \map_{\omega_{1}}x)$. 

Let us give a few simple examples of random maps, demonstrating
that even in the simplest cases the dynamics of random maps might
be rather nontrivial. For the sake of simplicity in all our
examples both here and in the sequel the collection of maps will 
consist of only two maps $\map_{1}, \map_{2}$ from the unit 
interval into itself and thus the corresponding distribution is 
described by the one parameter $p:=p_{1}$.

\begin{example}\label{ex:1} Pure contractive maps: $\map_{1}(x):= x/2$,
$\map_{2}(x):=x/2 + 1/2$. \end{example}%
According to \cite{Hut} (see also the next section) for each
value of the parameter $0<p<1$ the corresponding random map
$\rmap$ has the unique invariant measure, whose support
typically (when $p\ne1/2$) is a Cantor set.

\begin{example}\label{ex:2} A mixed case: $\map_{1}(x) := 1 - |2x-1|$,
$\map_{2}(x) := x/2$. \end{example}%
In distinction to the previous example, depending on the choice of 
the parameter $p$ the properties of the random map $\rmap$ differ 
qualitatively. Namely for $0\le p<\frac12$ the random map possesses the 
unique invariant measure concentrated at $0$. For $p>\frac12$ the 
invariant measure becomes non unique but there exists a unique 
absolutely continuous invariant one, whose density $h_{\rmap}$ is such 
that $h_{\rmap}|I_{k}=\gamma_k$ for any $k=1,2,\dots$ and intervals 
$I_{k}:=(2^{-(k-1)},2^{-k}]$. However the constants $\gamma_k$ are bounded 
on $k$ while $2/3<p<1/2$ and go to infinity with $k$ for $0\le p<2/3$.

The above examples are typical for statistical problems related
to random maps and in the next section we shall discuss separately 
properties of random maps expanding on average and contracting on 
average and thus we postpone the proof of the above claims till all 
needed technicalities will be introduced. It is worth note that our 
results related to the expanding on average random maps are very 
similar to those known for deterministic piecewise expanding maps 
\cite{Bl-mon} (see also results about the convergence to the 
Sinai-Bowen-Ruelle (\SBR) measure for multidimensional expanding 
on average random maps in \cite{KhKi, Buz}). Therefore in the 
corresponding sections we mainly demonstrate the similarities between 
these two types of systems. On the other hand, spectral analysis of 
the contracting on average random maps is completely new. In fact our 
interest in this type of systems is due to the analysis of spectral 
properties of Anosov maps in \cite{BKL}, where the presence of stable 
foliations relates the situation to our case. Some of the methods 
and ideas used in this paper were originated from the construction 
in \cite{BKL} needed to study the behavior of the transfer operators 
`along the stable foliation' and are based technically on the 
establishing of the so called Lasota-Yorke type inequalities. The 
nature of the system under study gives some advantages compare to 
the Anosov maps, in particular, the functional Banach spaces in our 
case do not depend on the finite structure of the map, the spectrum 
stability results are proven for a much broader class of random 
perturbations, and a more direct approximation scheme is elaborated.

The paper is organized as follows. In the following two sections
(2 and 3) we introduce the notions of contractive and expanding
on average random maps and discuss some of their basic properties
slightly generalizing known results about them. The main results
of the paper are described in section 4, where we construct and
study the Perron-Frobenius spectrum for random maps. We prove
also stochastic stability of isolated eigenvalues in this
spectrum and construct by means of a special random smoothing two
schemes of their finite rank operator approximation. To some extent 
these schemes generalize the well known Ulam approximation scheme
\cite{Ul} (see also discussion of its realization in 
\cite{Bl-mon,dellnitz,Fro,Ki4}). We discuss in detail why the
spectrum approximation by means of the original Ulam scheme is
not stable in the case of random maps (see also \cite{Bl-mon,BKL}). 
Since the original Ulam conjecture about the approximation of the \SBR 
measure (the leading element of the spectrum) still holds in our setting, 
in Lemma~\ref{l:counter-or-ulam} we construct the first (to the best 
of our knowledge) counterexample to this conjecture. Note however that 
the map in this counterexample is not only discontinuous, but the 
discontinuity occurs in a periodic turning point (compare to 
instability results for general random perturbations in \cite{Bl-mon}).

\section{Contracting on average random maps}\label{s:contract}

An important and well known case of random maps corresponds to
the situation when all the maps $\map_{i}$ are continuous and
their {\em contracting constants}
$$ \La_{\map_{i}} := \sup_{x,y\in X} \frac{|\map(x) - \map(y)|}{\rho(x,y)}
   \le \La < 1 .$$
Only the list of literature dedicated to various theoretical and
applied aspects of the theory of this class of random maps would
fill several pages, therefore we only give a reference to one of
the first (and giving a very essential contribution) works in
this field -- \cite{Hut}, to the monograph \cite{Fal}, and to one
of the recent publications \cite{FaLa}, where the reader can find
more references to recent results. Note that these papers are
dedicated mainly to questions related to dimensional and
multifractal properties of invariant sets and measures which we
shall not touch in this work.

We start the analysis of ergodic properties of these systems
using a weaker assumption, namely that the random map $\rmap$ is
contracting on average.

\begin{definition}\label{i:rmap-c} We shall say that the random map
$\rmap$ is {\em contracting on average} if its {\em contracting
constant}
$$ \Lambda_\rmap := \sum_i p_i \Lambda_{\map_i} < 1 .$$
\end{definition}

Let $\cM$ be the space of probability measures on $X$. Then the
Markov operator associated with the random map $\rmap$ for each
probabilistic measure $\mu \in\cM$ can be written as
$$ \rmap\mu := \sum_i p_i \mu \circ \map_i^{-1} .$$

Following the standard scenario we introduce in the space $\cM$ the
special metrics (often called Hutchinson metrics) \label{i:hatch} \cite{Hut}:
$$ \rho_H(\mu,\nu) := \sup\left\{\int h \, d\mu - \int h \, d\nu, \;
     h \in \IC^0(X,\IR), \; |h(x)-h(y)| \le \rho(x-y), \;
       \forall x,y \in X \right\} ,$$
where $\rho$ is the metrics on our original phase space $X$. It
is well known (see, for example, \cite{Hut}) that the pair
$(\cM,\rho_H)$ defines a compact metric space.

\begin{lemma} Let $h:X \to \IR$ be a continuous function and let
$\mu\in\cM$. Then
$$ \int h \, d (\rmap\mu) = \sum_i p_i \int h \circ \map_i \, d\mu .$$
\end{lemma}

\proof For each continuous function $h:X \to \IR$ there is a
sequence of piecewise constant approximating functions $h_n$
converging uniformly on $X$. Therefore%
\bea{\int h_n \, d (\rmap\mu)
  \a = \sum_i p_i \int h_n \, d(\mu \circ \map_i^{-1}) \\
  \a = \sum_i p_i \int_{\map_i X} h_n \, d(\mu \circ \map_i^{-1})
 = \sum_i p_i \int h_n \circ \map_i \, d\mu . }%
On the other hand, $\int h_n \, d (\rmap\mu) \to \int h \,
d(\rmap\mu)$ as $n \to \infty$, while for each pair $i,n$ the
function $h_n \circ \map_i$ is piecewise constant. Thus the
sequence of functions $\{h_n \circ \map_i\}_n$ converges
uniformly on $n$ to a function $h \circ \map_i$ which yields the
convergence
$$ \sum_i p_i \int h_n \circ \map_i \, d\mu \to
   \sum_i p_i \int h \circ \map_i \, d\mu .$$
\qed

The following result is a simple generalization of the well known
Hutchinson Theorem \cite{Hut}.

\begin{lemma}\label{un-erg1}
$\rho_H(\rmap\mu,\rmap\nu) \le \La_{\rmap}\cdot \rho_H(\mu,\nu)$
for any measures $\mu,\nu \in \cM$. In particular contraction on
average yields the strict ergodicity of the random map $\rmap$.
\end{lemma}

\proof Introduce the notation
$$ {\cal H} := \left\{h \in \IC^0(X,\IR), \; |h(x)-h(y)| \le \rho(x,y), \;
            \forall x,y \in X\right\} .$$
Then%
\bea{\rho(\rmap\mu,\rmap\nu)
 \a= \sup\lr{\int h\,d(\rmap\mu) - \int h \, d(\rmap\nu),\; h \in {\cal H}} \\
 \a= \sup\lr{\sum_i p_i \int h \circ \map_i \, d\mu
         - \sum_i p_i \int h \circ \map_i \, d\nu, \; h \in {\cal H}} .}%
Consider a function $\tilde h := \frac1\La_{\rmap} \sum_i p_i h
\circ \map_i$. For each pair of points $x,y \in X$ we have%
\bea{|\tilde h(x) - \tilde h(y)|
  \a\le \frac1{\La_{\rmap}} \sum_i p_i |h\circ \map_i (x) - h\circ\map_i(y)| \\
  \a\le \frac1{\La_{\rmap}} \sum_i p_i \rho(\map_i(x), \map_i(y))
    \le \frac1{\La_{\rmap}} \sum_i p_i \La_{\map_{i}}\cdot \rho(x,y)
    = \rho(x,y) .}%
Hence $\tilde h \in {\cal H}$. Introducing another set of functions
$$ \tilde {\cal H}
 := \lr{\tilde h \in {\cal H}: \; \exists h \in {\cal H}: \;
        \tilde h = \frac1{\La_{\rmap}} \sum_i p_i h \circ \map_i} ,$$
we can rewrite the distance between the images of the measures as
follows
$$ \rho_{H}(\rmap\mu,\rmap\nu)
 = \sup\lr{\frac1{\La_{\rmap}} \int \tilde h \, d\mu
                  - \frac1{\La_{\rmap}} \int \tilde h \, d\nu:
                         \; \tilde h \in \tilde {\cal H}} .$$
Now, since $\tilde {\cal H} \subset {\cal H}$, we come to the desired
estimate
$$ \rho_H(\rmap\mu,\rmap\nu) \le \La_{\rmap}\cdot \rho_H(\mu,\nu) ,$$
and thus the contraction on average yields the uniform
contraction in the space of measures. \qed

Note that the example~\ref{ex:1} satisfies the conditions of
Lemma~\ref{un-erg1}, while for the example~\ref{ex:2} the
conditions of Lemma~\ref{un-erg1} hold only when $0<p<1/3$.
Indeed,
$$ \sum_{i}p_{i}\La_{i} = 2p + \frac12(1-p)
                        = \frac32p + \frac12 < 1 .$$

On the first sight it seems that the continuity of the maps
$\map_i$ was not used in the proof, however it plays a very
important role in it. One can easily construct an example when
the absence of this property leads to the nonuniqueness of the
invariant measure.

\begin{example}\label{ex:4}
$\map_1(x) := \frac{x}2 \1{[0,1/2]}(x) + \frac{x+1}2 \1{(1/2,1]}(x)$,
$\map_2(x) := x$. \end{example}%
For each $p\in(0,1)$ the random map corresponding to
example~\ref{ex:4} possesses exactly two ergodic invariant
measures concentrated at points 0 and 1 respectively.

The contraction on average, that we assume in this section, does
not prevent some of the maps $\map_i$ to be expanding. Therefore,
despite the fact that some of the maps $\map_i$ may possess
several (not necessary a finite number) ergodic invariant
measures, the random map $\rmap$ under the assumptions of
Lemma~\ref{un-erg1} is strictly ergodic.

\section{Expanding on average random maps}

Results obtained in the previous section are based technically on
the contraction on average property. Now we are going to show that
the opposite assumption about the expansion on average leads to
the ideologically close result -- the existence of the absolutely
continuous invariant measure.

For the sake of simplicity we shall restrict ourselves here to the
analysis of piecewise $C^2$-smooth maps of the unit interval $[0,1]$
into itself with nondegenerate {\em expanding constants}
$$ \la_{\map_i} := \inf_x |\map_i'(x)| \ge \la > 0 .$$
It is straightforward to show that the transfer operator
corresponding to the random map $\rmap$ in the space $\IL_1$ can
be written as
$$ \IP_{\rmap} := \sum_i p_i \IP_i ,$$
where $\IP_i$ is the Perron-Frobenius operator, corresponding to
the map $\map_i$ and describing the dynamics of densities of measures
under its action (see a detailed discussion of properties of these
operators for example in \cite{Bl-mon}).

Denoting by $\Omega$ the set of values of the index $i$, we consider
for a given number $\La$ the sequence of sets
$$ \Omega^{(n)}(\La)
 := \{\omega \in \Omega: \; |(\rmap_\omega^n x)'| > \La^n
    \; {\rm for \ a. a.} \; x \in X \} ,$$
where
$$ \rmap_\omega^n x
 := \map_{\omega_n} \circ \map_{\omega_{n-1}} \circ \dots
                    \circ \map_{\omega_1} x $$
is the $n$-th point of a realization of a trajectory of the random map
$\rmap$ starting from the point $x$, and introduce the following
{\em regularity assumption}\label{i:reg-r}: there exist two constants
$\Lambda>1$ and $C<\infty$ such that%
\beq{reg-ass}{ \cP\{\Omega^{(n)}(\La) \} \ge 1 - Ce^{-\sqrt{n}} }%
for each positive integer $n$.

Denote by $\var(\cdot)$ the standard one-dimensional variation o
a function and by $\BV$ the space of functions of bounded variation
equipped with the norm $||\cdot||_{\BV}:=\var(\cdot) + ||\cdot||_{\IL_{1}}$.

The following result gives the decomposition for the transfer operator
under the considered assumptions.

\begin{theorem}\label{random-decomposition-exp}\cite{Mo,Bl-mon}
Let the regularity assumption (\ref{reg-ass}) holds. Then for
each pair of positive integers $n,k$ the following decomposition
takes place for the random map $\rmap$:
$$ \IP_\rmap^n = P_{n,k} + Q_{n,k} ,$$
and
\beq{r-exp-decom'}{ \var(P_{n,k} h)
    \le \const \left( \alpha^n\var(h) + \beta^k ||h|| \right) ,}
\beq{r-exp-decomp2}{||Q_{n,k}h|| < \const \sqrt{k}e^{-\sqrt{k}}
||h|| ,} for each function $h \in \BV$ and
$0<\alpha<1<\beta<\infty$. All constants above depend on the
choice of the map $\map_i$, but do not depend on $n$ and $k$.
\end{theorem}

One can find in the literature dedicated to the question of the
existence of absolutely continuous invariant measures in our
setting two types of sufficient conditions for this existence.
The first of these conditions obtained in \cite{Pe1} corresponds
to the strong expansion on
average \label{i:pel}%
\beq{Pel-ineq}{ \sum_{i} \frac{p_{i}}{\la_{\map_{i}}} < 1 ,}%
while the second, described in \cite{Mo}, is a weaker condition:%
\beq{Mor-ineq}{ \prod_{i} \la_{\map_{i}}^{p_{i}} > 1 .}%
One can easily show that the first of these assumption yields the
second one. On the other hand, as we shall show there is an
important difference between properties of invariant measures and
respectively random maps under these assumptions. To explain the
difference let us return to our regularity assumption and show
that it is even more general with respect the inequality
(\ref{Mor-ineq}). For this purpose we shall need the following
simple technical estimate.

\begin{lemma}\label{ineq-exp-random} Let $\{\xi_{i}\}_i$ be a
sequence of independent identically distributed (iid) random
variables having exponential moments up to some positive order
$s_0$, i.e. $\E{e^{s\xi}}<\infty$ for all $0<s<s_{0}$. Then for
each number $R>\E{\xi_{i}}$ there are constants $a<1,A<\infty$
such that
$$ \cP\{\frac1n \sum_{i=1}^{n}\xi_{i} > R\} < A a^{n} $$
for each positive $n$.
\end{lemma}

\proof By the exponential Chebyshev inequality
$$ \cP\{\frac1n\sum_{i=1}^{n}\xi_{i}>R\} \le e^{-sR} \E{e^{s\xi}} $$
for each positive number $s$. For our purpose it is enough to show
that the right hand side of this inequality decreases exponentially fast.
Note that for each number $x$ the following inequality holds
$$ |e^{x} - 1 - x| \le e^{|x|} - 1 - |x| .$$
Indeed, this is trivial for $x\ge0$ (since $e^x\ge+x$ and
$e^{-x}\le e^x$), while for $x<0$ we have $e^x\le+x$. Thus the
inequality can be reduced to
$$ -e^x + 1 + x \le e^{-x} -1 + x \qquad{\rm or}\qquad
   e^{-x}+e^{x}\ge2 ,$$
which is evidently correct. Therefore
$$ |e^{s\xi} - 1 - s\xi| \le e^{|s\xi|} - 1 - |s\xi| .$$
Assume first, that the values $\xi_{i}$ are bounded from below. Then
$$ \E{e^{|s\xi|}} - 1 - \E{|s\xi|} < \infty $$
for $|s|<s_{0}$, which yields the negativity of the left hand side of
the previous inequality. Therefore there are such positive constants
$s\in(0,\tilde s_{0})$ and $C$ that
$$ \E{e^{s\xi}} \le 1 + \E{\xi} + Cs^{2} \le e^{s\E{\xi} + Cs^{2}} $$
Thus, setting
$$ s=(R-\E{\xi})/(2C), \quad a=e^{-(R-\E{\xi}^{2})/(4C))} ,$$
we get the desired estimate. Observe that $s<\tilde s_{0}$ by the 
consruction. Therefore our inequalities make sense only if
$$ R\le R_{0}:=\E{\xi} + 2 C\tilde s_{0} .$$
This means that larger values of $R$ should be changed to $R_{0}$.
To finish the proof note that if the random values $\xi_{i}$ are not
bounded from below it is enough to `cut' them from below by means of
some constant and to apply the above argument to the result. \qed

\begin{lemma} The inequality (\ref{Mor-ineq}) implies the regularity
assumption (\ref{reg-ass}). \end{lemma}

\proof It is enough to apply Lemma~\ref{ineq-exp-random} to the
sequence of iid random values $\xi_{i}:=\ln\la_{\map_{i}}$. \qed

On the other hand, the following result shows that the opposite
statement does not hold.

\begin{example}\label{ex:5}
$$ \map_1(x)
 := \function{\frac34 - 2x        ,&\mbox{if }       0\le x<\frac14  \\
              \frac14 - \frac23 x ,&\mbox{if } \frac14\le x<\frac12  \\
              2 - 2x \mod1        ,&\mbox{otherwise} ,} $$
while $\map_2(x) := \map_1(x) + 1/12 \mod1$. Graphs of these maps
are shown on Fig.~\ref{reg-mor-counter}.
\end{example}
\Bfig(150,150)
      {\bline(0,0)(1,0)(150)   \bline(0,0)(0,1)(150)
       \bline(0,150)(1,0)(150) \bline(150,0)(0,1)(150)
       \bline(0,0)(1,1)(150)
       \thicklines
       \bline(0,114)(1,-2)(38)  \bline(38,38.5)(3,-2)(37)
       \bline(75,14)(1,2)(68)   \bline(142,0)(1,2)(8)
       \bline(0,100)(1,-2)(37)  \bline(38,25.5)(3,-2)(37)
                                \bline(75,0)(1,2)(75)
       \thinlines
       \bline(38,38.5)(0,-1)(42) \bline(75,14)(0,-1)(17)
       \put(35,-10){$\frac14$}   \put(73,-10){$\frac12$}
       \put(-10,112){$\frac34$}  \put(-10,96){$\frac23$}
       \put(23,75){$\map_1$}     \put(15,35){$\map_2$}
      }{Example when the regularity condition holds while the
        condition~(\ref{Mor-ineq}) breaks down \label{reg-mor-counter}}

\begin{lemma} The random map $\rmap$ in the example~\ref{ex:5} for any
$0<p<1$ satisfies the regularity assumption, while the
condition~(\ref{Mor-ineq}) breaks down. It is interesting that in
this example the stronger assumption~(\ref{Pel-ineq}) holds for
the second iterate $\rmap^2$.
\end{lemma}

\proof Both maps $\map_{i}$ are piecewise linear and the moduli
of their derivatives take only two values $2$ (on the intervals
$(0,\frac14)$ and $(\frac12,1)$) and $\frac23$ (on the interval
$(\frac12,1)$). Since both these maps transform the interval
$(\frac14,\frac12)$ into $(0,\frac14)$ and on the remaining
interval $(\frac12,1)$ the derivatives of both maps is equal to
$2$, it follows that for any $i,j\in\{1,2\}$ the inequality
$$ |(\map_{i}\map_{j}x)'| \ge 2 \cdot \frac23 = \frac43 > 1 $$
holds. Thus we have checked the regularity assumption. Now observe
that the derivatives of both maps is strictly less than $1$ on the
interval $x\in(\frac14,\frac12)$, which contradicts to the
condition~(\ref{Mor-ineq}).

It remains to check the condition~(\ref{Pel-ineq}) for the second
iterate $\rmap^2$, which turns out to be a consequence of the
fact that expanding constants for the maps $(\map_{i}\map_{j}x)$
are not less than $\frac43>1$ (according to the inequality
above). \qed

Note that since the right hand side of the
inequality~(\ref{r-exp-decom'}) contains the term $\beta^k$ with
$\beta>1$ the Theorem~\ref{random-decomposition-exp} guarantees
only estimates of the type
$$ \var(P_{n,k} \overline h) \le \const \frac{\beta^k}{1-\alpha} ,$$
which means that despite the fact that the density of the invariant
measure is integrable, it might be not a function of bounded variation.
In fact, the example 2 from Section~\ref{sec:intro} demonstrate this
phenomenon when the parameter $p$ belongs to $(\frac12,\frac23)$.
Moreover, in this example the density not only not a function of bounded
variation, but it goes to infinity in the vicinity of the origin.

It turns out that under a stronger assumption (\ref{Pel-ineq})
the standard Lasota-Yorke inequality is valid for the random map
and thus the invariant density is a function of bounded variation.

\begin{theorem}\cite{Pe1}\label{rand-map-bounded}
Let for a.a. $x\in [0,1]$ the inequality
$$ \sum_i \frac{p_i}{|\map'_i(x)|} \le \gamma < 1 $$
holds. Then there are constants $C,\beta<\infty$ such that for each
$n\in\IZ_{+}$ and a function $h\in\BV$ the Lasota-Yorke inequality holds:%
\beq{rand-map-LY}{ \var(\IP_{\rmap}^{n}h)
                   \le C\gamma^{n}\var(h) + \beta||h|| ,}%
from where (as usual) it follows that for each nonnegative
function $h\in\IL^1$ with $||h||=1$ the limit
$$ \lim_{n\to\infty} \frac1n \sum_{k=0}^{n-1} \IP_{\rmap}^k h
 =: h_{\rmap} ,$$
exists and
$$ \IP_{\rmap} h_{\rmap} = h_{\rmap}, \qquad
   \var(h_{\rmap}) \le \const ,$$
while which respect to the absolutely continuous invariant measure
$\mu_{\rmap}$ with the density $h_{\rmap}$ the correlations decay
exponentially.
\end{theorem}

Now we are able to finish the analysis of the example~2 from
Section~\ref{sec:intro}. Observe that for $p>\frac12$ we have
$$ \prod_{k} \la_{\map_{k}}^{p_{k}}
 = 2^{p}\cdot \left(\frac12\right)^{1-p} = 2^{2p-1} > 1 ,$$
which yields the condition~(\ref{Mor-ineq}) and, hence, our
regularity assumption. Therefore for $p>1/2$ there exists an
absolutely continuous $\rmap$-invariant measure. On the other
hand, for $\frac23<p<1$ the condition~(\ref{Pel-ineq}):
$$ \sum_{k}\frac{p_{k}}{|\map'_{k}(x)|}
 = \frac{p}2 + 2(1-p) = 2 - \frac32p < 1 $$
holds for a.a. $x\in[0,1]$. Thus we can apply
Theorem~\ref{rand-map-bounded}, whereis the boundedness of the
density of the invariant measures follows.

\section{Perron-Frobenius spectrum (PF-spectrum)}
\label{sec:spectr}

In the previous sections we have restricted the analysis of
statistical features of random maps to the properties of their
invariant measures and more specifically Sinai-Bowen-Ruelle (\SBR
measures). From a more general point of view the SBR measure is
the eigenfunction of the Perron-Frobenius (transfer) operator of
our random map in a suitable Banach space corresponding to the
leading eigenvalue (1). Therefore it is very natural to extend
the analysis of the dynamics to the complete spectrum of this
operator.

It is worth note that the interest to the PF-spectrum is based to
a large extent on the fact that the subleading elements of the
spectrum define the rate of mixing (convergence to the \SBR
measure, correlation decay, etc.).

\subsection{Definition of the PF-spectrum}

Let us start with the short description of objects related to the
notion of the spectrum which we shall need further. Let $\IP: \cB
\to \cB$ be a bounded linear operator in a complex Banach space
$(\cB, \nB{\cdot})$. As usual we denote by $\sp_\cB(\IP)$ its
{\em spectrum}\label{i:sp}, which is defined as a complement to
the set of regular elements, i.e. to the points $z\in\IC$ such
that {\em the resolvent}\label{i:res} $(zI - \IP)^{-1}$ of the
operator is defined in the entire space and hence is bounded. The
maximal (on modulus) element of the spectrum is called the {\em
spectral radius}\label{i:sp-r}:
$$ \rho_\cB(\IP) := \sup\{|z|: \; z \in \sp_\cB(\IP) \} . $$
As it is well known \cite{DS} the spectral radius can be
calculated by the following formula:
$$ \rho_\cB(\IP) = \lim_{n \to \infty} \nB{\IP^n}^{1/n} .$$
Browder \cite{Bro} introduced the notion of {\em essential spectrum}
$\esssp_\cB(\IP)$ of a bounded linear operator $\IP$ as a union of
the elements of the spectrum $z \in \sp_\cB(\IP)$ such that at least
one of the following properties holds:
\begin{enumerate}
   \item The region of values of the operator $zI-\IP$ is not bounded in $\cB$.
   \item $\cup_{n \ge 0} {\rm ker}((zI - \IP)^n)$ is infinite dimensional.
   \item $z$ is a limit point of the spectrum $\sp_\cB(\IP)$.
\end{enumerate}
Outside of the essential spectrum only a countable number of {\em
isolated}\label{i:sp-iz} eigenvalues of the operator $\IP$ may
occur. Naturally the {\em essential}\label{i:sp-r-ess} spectral
radius $\rho_{\cB,{\rm ess}}(\IP)$ of the operator $\IP$ is
defined as the minimal nonnegative number such that all elements
of the spectrum $\sp_\cB(\IP)$ outside of the disc $\{z \in \IC:
\; |z| \le \rho_{\cB,{\rm ess}}(\IP) \}$ are isolated eigenvalues
of finite multiplicity. It turns out \cite{Nu} that the essential
spectral radius can be calculated by a formula similar to the
one for the usual spectral radius:%
\beq{ess-sp-rad}{\rho_{\cB,{\rm ess}}(\IP)
          = \lim_{n\to\infty}\norm{\IP^n}{\cB,{\rm ess}}^{1/n} ,}
but for the specialized seminorm:
$$ \norm{\IP}{\cB,{\rm ess}}:=\inf\{\nB{\IP-K}:\;K:\cB\to\cB\
          \hbox{   -- a compact operator}\} .$$
Clearly for each $\ep>0$ the set
$$ \sp(\IP) \cap \{z\in\IC: \; |z|\ge \rho_{\cB,{\rm ess}}(\IP) + \ep\} $$
consists of a finite number eigenvalues
$\la_1,\la_2,\dots,\la_{M(\ep)}$ of the operator $\IP$.
Schematically on the complex plane the spectrum can be
represented as a disk of radius $\rho_{\cB,{\rm ess}}(\IP)$
centered at the origin (describing the essential spectrum) and
(not more than countable) collection of points between this disk
and the circle of radius $\rho_\cB(\IP)$ also centered at the
origin.

\subsection{Spectrum for the case of contracting on average random maps}

In Section~\ref{s:contract} it was shown that a contracting on
average random map $\rmap$ possesses the only one invariant
measure $\mu_{\rmap}$ to which the sequence of iterations
$\{\rmap^{n}\mu\}$ converges (exponentially fast in the
Hutchinson metrics) for each probabilistic initial measure $\mu$.
From the point of view of dynamical system theory the following
question after this is the analysis of the spectrum of this
convergence. The problem here is that typically the limit measure
$\mu_{\rmap}$ is not absolutely continuous, which rules out the
description of the transfer operator $\IP_{\rmap}$ in a space of
a reasonably `good' functions, for example, in the space of
functions of bounded variation. To overcome this difficulty we
shall study the action of the operator $\IP_{\rmap}$ in a much
larger space of generalized functions equipped with a norm
induced by the Hutchinson metrics.

Let $(X,\rho)$ be a $d$-dimensional smooth manifold with a finite
collection of continuous maps $\{\map_{i}\}$ having bounded
Lipschitz constants $\La_{\map_{i}}$ from $X$ into itself, and a
collection of probabilities $\{p_{i}\}$, defining the random map
$\rmap$. Remind that the contraction on average means that
$$ \La_{\rmap} := \sum_{i} p_{i} \La_{\map_{i}} < 1 .$$

Before to define our space of generalized functions we need first
to define the class of test-functions $\phi:X\to\IR^{1}$. For
this purpose we introduce the following functionals:%
\bea{H_{\alpha}(\phi) \a:= \sup_{x,y\in X,\; 
       \rho(x,y)\le\nu} \frac{|\phi(x) - \phi(y)|}{\rho^{\alpha}(x,y)} ,\\
     V_{\alpha}(\phi) \a:= H_{\alpha}(\phi) + |\phi|_{\infty} ,}%
where $|\phi|_{\infty}:={\rm esssup}|\phi|$, and the constant
$\nu\in(0,1]$. The first of these functionals is the H\"older
constant with the exponent $\alpha<1$, while the second one for
each finite nonnegative value of the parameter $a$ is the norm in
the Banach space of $\alpha$-H\"older functions on $X$, which we
shall denote by $\IC^{\alpha}$. Without the loss of generality we
shall assume that the diameter of the phase space $\sup_{x,y\in
X}\rho(x,y)\le1$. Note that the another restriction
$\rho(x,y)\le\nu$ is introduced only to be able to work with the
exponential map on general smooth manifolds: in the case of a
flat torus this restriction can be omitted (or simply one can set
$\nu=1$).

Consider now the space of generalized functions $\cF$ on $X$ with
the norm defined in terms of the test-functions from the space
$\IC^{\alpha}$:
$$ \na{h}{\alpha} := \sup_{V_{\alpha}(\phi)\le1} \int h\phi .$$
The proof that the functional $\na{\cdot}{\alpha}$ is indeed a norm in
this space is standard and we leave it for the reader.

Denote by $\cF_{\alpha}$ the closure of the set of bounded in the
norm $\na{\cdot}{\alpha}$ generalized functions from $\cF$.

\begin{lemma}\label{l:holder} $\cF_{\beta} \subseteq\cF_{\alpha}$
for any numbers $0<\alpha\le\beta\le1$ and each function $\phi\in\IC^\beta$
the inequality $V_\beta(\phi) \le V_\alpha(\phi)$ holds.
\end{lemma}

\proof Indeed,
$$ \frac{|\phi(x) - \phi(y)|}{\rho^{\beta}(x,y)}
 = \frac{|\phi(x) - \phi(y)|}{\rho^{\alpha}(x,y)}
   \rho^{\beta-\alpha}(x,y) \le H_{\alpha}(\phi) .$$
\qed

Extending the standard definition of the Perron-Frobeniusa
operator to the action in the space of generalized functions we
get the representation
$$ \int \IP_{\rmap} h \cdot \phi
 = \int h \cdot \sum_{i} p_i \, \phi \circ \map_{i}
 =: \int h \cdot (\phi\circ\rmap) $$
for each test-function from $\phi\in\IC^{\alpha}$.

Let us fix some constants $0<\alpha<\beta\le1$, $0<a<\infty$, whose
exact values we shall define later. Our first aim is to derive
a version of the Lasota-Yorke inequality for the action in the space
$\cF_{\alpha}$. For $q\in[0,1]$ define a function
$$ \La_{\rmap}(q) := \sum_{i}p_{i}\La_{\map_{i}}^{q} ,$$
which we shall need in this derivation.

Introduce additionally the notation ${\overline G}=\{G_i,p_i\}$
for the random map defined by the collection of maps $G_i$ and
the distribution $\{p_i\}$.

\begin{lemma}\label{l:holder-sup} The superposition of any pair of
random maps ${\overline G}=\{G_i,p_i\}$ and ${\overline
G'}=\{G'_i,p'_i\}$ acting on the same manifold $(X,\rho)$ and
satisfying the Lipschitz condition is again the random map
${\overline G'}\circ{\overline G} := \{G'_i\circ G_j,p'_ip_j\}$
and $\La_{{\overline G'}\circ{\overline G}}(q)
 \le \La_{{\overline G'}}(q) \cdot \La_{{\overline G}}(q)$
for each $q\in[0,1]$.
\end{lemma}

\proof Indeed,%
\bea{\La_{{\overline G'}\circ{\overline G}}(q)
 \a= \sum_i \sum_j p'_i p_j \La_{G'_i\circ G_j}^q
   \le \sum_i \sum_j p'_i p_j \La_{G'_i}^q \cdot \La_{G_j}^q \\
 \a= \left( \sum_i p'_i \La_{G'_i}^q \right) \cdot
     \left( \sum_j p_j \La_{G_j}^q \right)
   = \La_{{\overline G'}}(q) \cdot \La_{{\overline G}}(q) .}%
\qed

\begin{theorem}\label{th-LY-rand-contr} For each number $\kappa>2$
and for any $h\in\cF_{\alpha}$ and $n\in\IZ_+$
the Lasota-Yorke inequality holds: %
\beq{LY-random-ifs}{
   \na{\IP_{\rmap}^nh}{\alpha}
   \le \kappa\La_\rmap^n(\alpha) \na{h}{\alpha}
     + \const\cdot (\kappa-2)^{-1/\alpha} \na{h}{\beta} .}
\end{theorem}

\proof We start from the proof of the following two inequalities:%
\beq{LY-random-ifs-1}{ \na{\IP_{\rmap}h}{\beta} \le \na{h}{\beta} ,}%
\beq{LY-random-ifs-2}{
   \na{\IP_{\rmap}h}{\alpha}
   \le \kappa\La_\rmap(\alpha) \na{h}{\alpha}
     + \const\cdot (\kappa-2)^{-1/\alpha}  \na{h}{\beta} .}%
By the definition of the H\"older constant we have:
$$ \frac{|\phi(\map_i x) - \phi(\map_i y)|}{\rho^\alpha(x,y)}
 = \frac{\rho^\alpha(\map_i x, \map_i y)}{\rho^\alpha(x,y)}
   \frac{|\phi(\map_i x) - \phi(\map_i y)|}{\rho^\alpha(\map_i x, \map_i y)}
 \le \La_{\map_i}^\alpha H_\alpha(\phi) .$$
Hence,
$$ \frac{\sum_i p_i |\phi(\map_i x) - \phi(\map_i y)|}{\rho^\alpha(x,y)}
 \le \sum_i p_i \La_{\map_i}^\alpha H_\alpha(\phi) .$$
Thus,
$$ H_\alpha\left(\sum_i p_i (\phi\circ\map_i)\right)
 \le \La_\rmap(\alpha) H_\alpha(\phi) < H_\alpha(\phi) .$$
On the other hand, since
$$  |\phi\circ\map_i|_\infty \le |\phi|_\infty ,$$
then
$$ \sum_i p_i |\phi\circ\map_i|_\infty \le |\phi|_\infty .$$
Therefore for each $\beta\in(0,1]$ we have
$$ V_\beta\left(\sum_i p_i |\phi\circ\map_i|\right) \le V_\beta(\phi) ,$$
which implies the inequality~(\ref{LY-random-ifs-1})
for each $\beta\in(0,1]$ and a function $h\in\cF_\beta$.

To prove the second inequality~(\ref{LY-random-ifs-2}) we need
more delicate estimates.

Introduce the following notation:
$$ B_\delta(x) := \{y\in X:\; \rho(x,y)\le\delta \} \qquad
   B_\delta    := \{\xi\in {\cal T}_x X: \; |\xi|\le\delta\} ,$$
i.e. $B_\delta(x)$ is the ball of radius $\delta$ centered at the
point $x$ in the space $X$, while $B_\delta$ is the ball of radius
$\delta$ centered at the origin in the tangent space ${\cal T}_x X$.
For each point $x\in X$ consider the exponential map
$$ \Psi_x := \exp_x : B_\delta \subseteq {\cal T}_x X \to X .$$
Choosing $\delta>0$ small enough we always can assume that
$\Psi_x B_\delta \subset B_\nu(x)$.

Denoting by $m$ the Lebesgue measure on $X$, we introduce the
following smoothing operator
$$ Q_\delta \phi(x)
 := \frac{\int_{\Psi_x B_\delta} \phi(y) m(dy)}{m(\Psi_x B_\delta)}
  = \frac{\int_{B_\delta} \phi(\Psi_x z)\cdot J\Psi_x(z) \, dz}
         {\int_{B_\delta} J\Psi_x(z) \, dz} ,$$
where $J\Psi_x$ is the Jacobian of the map $\Psi_x$.

Let us estimate the H\"older constant of the function $Q_\delta \phi$:%
\bea{|Q_\delta \phi(x) - Q_\delta \phi(y)|
 \a\le \left| \int_{B_\delta} J\Psi_x(z) \, dz
          - \int_{B_\delta} J\Psi_y(z) \, dz \right| \cdot
     \left| \frac{\int_{B_\delta} \phi(\Psi_y z)\cdot J\Psi_y(z) \, dz}
                 {\int_{B_\delta} J\Psi_x(z) \, dz \cdot
                  \int_{B_\delta} J\Psi_y(z) \, dz} \right| \\
 \a~ + \frac1{\int_{B_\delta} J\Psi_x(z) \, dz} \cdot
     \left| \int_{B_\delta} \phi(\Psi_x z)\cdot J\Psi_x(z) \, dz
          - \int_{B_\delta} \phi(\Psi_y z)\cdot J\Psi_y(z) \, dz \right| \\
 \a\le 2\left( \sup_x J\Psi_x \right)^2 \cdot |\phi|_\infty \cdot
     \frac{|B_\delta\setminus\Psi_x^{-1}\Psi_y B_\delta|
          + |\Psi_x^{-1}\Psi_y B_\delta \setminus B_\delta|}{|B_\delta|} \\
 \a\le \frac1\delta C_1 |\phi|_\infty \rho(x,y) ,}%
where the constant $C_1$ depends only on the properties of the manifold $X$.

Now we estimate how much the operator $Q_\delta$ differs from the
identical operator:
$$  |Q_\delta\phi(x) - \phi(x)|
 \le \frac{\int_{B_\delta} |\phi(\Psi_x(z)) - \phi(x)| \cdot J\Psi_x(z) \, dz}
          {\int_{B_\delta} J\Psi_x(z) \, dz}
 \le C_2 \delta^\alpha H_\alpha(\phi) ,$$
where the constant $C_2$ also depends only on the properties of the
manifold $X$.

On the other hand,%
\bea{\a H_\alpha\left(\sum_i p_i (\phi\circ\map_i)
            - Q_\delta\left(\sum_i p_i (\phi\circ\map_i)\right)\right) \\
 \a~\le H_\alpha\left(\sum_i p_i (\phi\circ\map_i)\right)
   + H_\alpha\left(Q_\delta\left(\sum_i p_i (\phi\circ\map_i)\right)\right) \\
 \a~\le 2\La_\rmap(\alpha) H_\alpha(\phi) .}%
Thus,
$$ V_\alpha\left(\sum_i p_i (\phi\circ\map_i)
            - Q_\delta\left(\sum_i p_i (\phi\circ\map_i)\right)\right) \CR
 \le 2\La_\rmap(\alpha) H_\alpha(\phi)
   + C_2 \delta^\alpha H_\alpha(\phi) .$$
Gathering the obtained estimates and using Lemma~\ref{l:holder} we come to%
\bea{\na{\IP_\rmap h}{\alpha}
 \a\le \sup_{V_\alpha(\phi) \le 1}
         \int h \cdot Q_\delta\left(\sum_i p_i (\phi\circ\map_i)\right) 
   + \sup_{V_\alpha(\phi) \le 1}
         \int h \cdot \sum_i p_i \left(\phi\circ\map_i
                           - Q_\delta(\phi\circ\map_i)\right) \\
 \a\le \sup_{V_\beta(\phi) \le 1}
           V_\alpha\left(Q_\delta\left(\sum_i p_i
                   (\phi\circ\map_i)\right)\right)
           \na{h}{\beta} \\
   \a~ + \sup_{V_\alpha(\phi) \le 1}
           V_\alpha\left( \sum_i p_i \left(\phi\circ\map_i
                           - Q_\delta(\phi\circ\map_i)\right) )\right)
           \na{h}{\alpha} \\
 \a\le \frac{C_1}\delta \na{h}{\beta}
   + \left( 2\La_\rmap(\alpha) + C_2\delta^\alpha \right) \na{h}{\alpha} .}%
Therefore choosing the value of the parameter $\delta$ such small that
$C_2\delta^\alpha = \kappa-2$, we get the inequality~(\ref{LY-random-ifs-2}).

Now according to Lemma~\ref{l:holder-sup} and above inequalities we get%
\bea{\na{\IP_{\rmap}^n h}{\alpha} 
  \a= \na{\IP_{\rmap^n} h}{\alpha} 
   \le \kappa\La_{\rmap^n}(\alpha) \na{h}{\alpha}
     + \const\delta^{-1} \na{h}{\beta} \\
  \a\le \kappa\La_{\rmap}^n(\alpha) \na{h}{\alpha}
     + \const\cdot (\kappa-2)^{-1/\alpha} \na{h}{\beta} ,}%
which finishes the proof of the inequality~(\ref{LY-random-ifs}). \qed

\begin{lemma}\label{l:convex-unique}
The function $\La_{\rmap}(\cdot)$ is convex, takes values
strictly less than $1$ in the interval $(0,1]$, and under the condition%
\beq{ne:uniq-rand}{\prod_i \La_{\map_{i}}^{p_i \La_{\map_{i}}} < 1}%
its unique point of minima either lies inside of this interval, or
is larger than $1$ otherwise.
\end{lemma}

\proof By the definition of the contraction on average we have
$\La_{\rmap}(1)=\La_{\rmap}<1$. On the other hand, $\La_{\rmap}(0)=1$,
and
$$ \frac{d^{2}}{dq^{2}} \La_{\rmap}(q)
 = \sum_{i}p_{i} \La_{\map_{i}}^{q}
                 \cdot \left(\ln \La_{\map_{i}} \right)^{2} > 0 .$$
Thus the function $\La_{\rmap}(q)$ is strictly convex and for
any $q\in(0,1]$ it is less than $1$. Let us show that this function
strictly decreases at $0$. Indeed,
$$ \frac{d}{dq} \La_{\rmap}(0) = \sum_{i} p_{i} \ln\La_{\map_{i}} .$$
On the other hand, from the contraction on average
$$ \sum_{i} p_i \La_{\map_{i}} < 1 $$
using the convexity of the logarithmic function, we get:
$$ \sum_{i} p_{i} \ln\La_{\map_{i}} < \ln1 =0  ,$$
which proves that $\frac{d}{dq} \La_{\rmap}(0) < 0$.
It remains to check the last statement.
$$ \frac{d}{dq} \La_{\rmap}(1)
 = \sum_i p_i \La_{\map_{i}}\cdot \ln\La_{\map_{i}}
 = \sum_i \ln\La_{\map_{i}}^{p_i \La_{\map_{i}}}
 = \ln\left( \prod_i \La_{\map_{i}}^{p_i \La_{\map_{i}}} \right) .$$
Thus due to the inequality~(\ref{ne:uniq-rand}) the unique (due to the
strict convexity of the function $\La_\rmap(\cdot)$) solution of the
equation $\frac{d}{dq} \La_{\rmap}(q)=0$ (the point of minima of the
function $\La_\rmap(\cdot)$) belongs to the interval $(0,1)$. On the
other hand, if the inequality~(\ref{ne:uniq-rand}) does not hold this
solution is greater than $1$.
\qed

A typical behavior of the function $\La_{\rmap}(q)$ is shown
Fig.~\ref{convex-q}.
\Bfig(150,150)
      {\bline(0,0)(1,0)(150)   \bline(0,0)(0,1)(150)
       \bline(0,150)(1,0)(150) \bline(150,0)(0,1)(150)
       \bline(20,20)(1,0)(110) \bline(20,20)(0,1)(110)
       \bline(120,20)(0,1)(70) \bline(86,20)(0,1)(47)
       \bezier{500}(20,120)(80,30)(120,90)
       \put(15,15){$0$} \put(135,20){$q$} \put(15,135){$\La_{\rmap}(q)$}
       \put(13,117){$1$} \put(117,10){$1$} \put(83,10){$\tilde q$}
      }{A typical behavior of the function $\La_{\rmap}(q)$. \label{convex-q}}
Denote by $\tilde q$ the value of the parameter $q$ corresponding
to the unique (due to the strict convexity) minima of the
function $\La_{\rmap}(q)$. The value $\tilde q$ is positive since
$\frac{d}{dq} \La_{\rmap}(0)<0$. The position of $\tilde q$ with
respect to $1$ is defined by the sign of the derivative of the
function $\La_{\rmap}(q)$ at $1$ which might be both positive and
negative. For example, if all the maps are contractive then this
sign is negative and the function $\La_{\rmap}(q)$ strictly
decreases on the interval $[0,1]$. On the other hand, the map in the 
example~\ref{ex:2} satisfies the condition~(\ref{ne:uniq-rand}) 
if $0<p<1/5$ and in this case $\tilde q=\frac12\log_2(1/p-1)$.
We describe the properties of the function $\La_{\rmap}(q)$ in
such detail because the fact that it can grow in the vicinity of
the point $1$ plays an important role in further calculations.

\begin{lemma} The unit disk in the strong norm $\na{\cdot}{\alpha}$
is a compact set in the weak norm $\na{\cdot}{\beta}$
\end{lemma}

\n The {\bf proof} of this statement follows immediately from
standard results on the enclosure of the spaces of H\"older
functions.

Above statements imply by the Ionescu-Tulchea and Marinescu
Theorem \cite{ITM} the quasicompactness of the operator
$\IP_\rmap$, and the validity of the following based on Nussbaum
Theorem \cite{Nu} estimate of its essential spectral radius.

\begin{lemma}\label{l:hennion}\cite{Hen} The essential spectal radius
of the operator $\IP_\rmap:\cF_\alpha\to\cF_\alpha$ belongs to the
disk or radius $\La_{\rmap}(\alpha)$ centered at zero. \end{lemma}

Note that the estimates leading to the Lasota-Yorke type inequalities 
depend sensitively on the choice of the value of the parameter $\alpha$. 
Thus it is reasonable to choose the value of $\alpha$ which yields 
the smallest (and hence the best) available estimate of the essential 
spectral radius. Normally (compare to \cite{BKL}) this value is equal 
to $1$ which is unavailable since we consider only H\"older continuous 
test functions. However in our case Lemma~\ref{l:convex-unique} shows 
that under the condition~(\ref{ne:uniq-rand}) the optimal value of 
$\alpha$ may be strictly less than $1$.

An immediate corollary to Lemma~\ref{l:hennion} is the existence
of a constant $\gamma\in[\La_{\rmap}(\alpha),1)$ such that the
set $\sp(\IP_\rmap)\setminus\{|z|\leq\gamma\}$ consists of a
finite number of periferal eigenvalues $r_1,\dots,r_N$ of finite
multiplicity. Denote by $P_1,\dots,P_N$ the corresponding
spectral projectors and set $P:=1_{\cF_\alpha} - \sum_{j=1}^N
P_j$. Then the rank$(P_j)<\infty$, $\IP_\rmap P_j = r_j P_j$
$(j=1,\dots,N)$, and the spectral radius of the operator
$\IP_\rmap P$ does not exceed $\gamma$.

Besides,
\begin{itemize}
\item If $|r|=1$ the operator
 \beq{eq:P_lambda}{P_r
   := \lim_{n\to\infty} \frac1n \sum_{k=0}^{n-1} r^{-k}\IP_\rmap^k
    = \sum_{j=1}^N \lim_{n\to\infty} \frac1n \sum_{k=0}^{n-1}
          \left(\frac{r_j}r\right)^k P_j
    =\function{P_j, &\mbox{if } r=r_j \\
                 0, &\mbox{otherwise}} }
 is well defined in the $\na{\cdot}{\alpha}$-norm.
 In particular, $P_j=P_{r_j}$, and since
 $\int|P_{r_j}f| \leq \int|f|$ for all $f\in\IC^1(X,\IR^1)$, then
 operators $P_j$ can be extended continuously to the entire space $\IL^1$.

\item For any function $f\in P_j\cF_\alpha$ there is a finite Borel signed
 measure $\mu_f$ on $X$ such that $\langle f,\phi\rangle = \int\phi\,d\mu_f$
 for all $\phi\in\IC^1(X,\IR^1)$.
 $r_1:=1\in\sp(\IP_\rmap)$, $\mu:=\mu_{P_1 1}$ -- a positive measure,
 $\mu(X)=m(X)$, and all signed measures $\mu_f$ are absolutely continuous
 with respect to $\mu$.

 One can interpret these statements as follows: for
 $f,\phi\in\IC^1(X,\IR^1)$ and $|r|=1$,%
  \beq{eq:Plambda}{ \langle P_r f,\phi\rangle
    = \lim_{n\to\infty}
      \frac1n \sum_{k=0}^{n-1} r^{-k}\langle\IP_\rmap^k f,\phi\rangle
    = \lim_{n\to\infty}
      \frac1n \sum_{k=0}^{n-1} r^{-k}\int\phi\circ\rmap^k\cdot f .}%
 Hence,
 $\langle P_r f,\phi\rangle\leq|\phi|_\infty\cdot\int|f|$, so
 $P_r f$ can be extended by continuity to a continuous linear functional
 on $\IC^0(X,\IR^1)$, and by Riss Theorem there is a measure $\mu_{P_r f}$,
 such that $\langle P_r f, \phi\rangle = \int\phi\,d\mu_{P_r f}$.
 If $r=1$ and $f,\phi\geq0$ then
 $\int\phi\,d\mu_{P_1f} = \langle P_1f,\phi\rangle\geq0$
 and $\langle P_1f,1\rangle = \int f$ according to (\ref{eq:Plambda}).
 $\mu_{P_11}$ is a positive measure and $r=1$ is an eigenvalue of the
 operator $\IP_\rmap$. Finally, it follows from (\ref{eq:Plambda}) that
 for any $\phi\geq0$ we have
  \begin{displaymath}
    |\langle P_jf,\phi\rangle|
    \leq \lim_{\n\to\infty}
         \frac1n \sum_{k=0}^{n-1}\int\phi\circ\rmap^k\cdot |f|
    \leq |f|_\infty\,\langle P_11,\phi\rangle
    = |f|_\infty\,\int\phi .
  \end{displaymath}
 Besides, $\mu_{P_jf}$ is absolutely continuous with respect to $\mu$.

 It remains to show that $P_j\cF_\alpha = V_j := P_j(\IC^1(X,\IR^1))$.
 Since $V_j\subseteq P_j\cF_\alpha$ and they are finite dimensional
 linear subspaces in $\cF_\alpha$, then this statement immediately follows
 from the denseness of the space $\IC^1(X,\IR^1)$ in $\cF_\alpha$.

\item $\rmap^*\mu=\mu$, since
  $\int\phi\,d(\rmap^*\mu)=\int\phi\circ \rmap\,d\mu
   =\lim_{n\to\infty} \frac1n \sum_{k=0}^{n-1} \int\phi\circ \rmap^{k+1}
   =\int\phi\,d\mu$ for all $\phi\in\IC^1(X,\IR^1)$.

\item If $r=1$ is a simple eigenvalue and there are no other eigenvalues
 equal to $1$ on modulus, then $P_1f=\langle f,1\rangle\cdot P_11$
 for all $f\in\cF_\alpha$. This follows immediately from the fact that
 $\langle P_1f,1\rangle = \langle f,1\rangle$ (see (\ref{eq:Plambda}).

\item $\mu$ is a \SBR measure, since for each $\phi\in\IC^1(X,\IR^1)$
  \begin{displaymath}
    \lim_{n\to\infty}
       \int\phi\,d\bigg(\frac1n \sum_{k=0}^{n-1}\rmap^{*k}m\bigg)
    = \lim_{n\to\infty}  \frac1n \sum_{k=0}^{n-1}\int\phi\circ\rmap^k
    = \langle P_11,\phi\rangle
    = \int\phi\,d\mu\ .
  \end{displaymath}
\end{itemize}

So far the only isolated eigenvalue that we were able to
identify explicitly was the unit eigenvalue (the leading one).
Let us discuss the following argument. According to 
Lemma~\ref{l:hennion} the essential spectral radius of the operator 
$\IP_\rmap$ is not larger than $\La_\rmap(\alpha)$. On the other 
hand, the Lemma~\ref{un-erg1} garantees the convergence to the limit 
measure with the rate at least $1>\La_\rmap>\La_\rmap(\alpha)$ 
(by Lemma~\ref{l:convex-unique}). Thus if there is a function 
$f\in\cF_\alpha$ such that $\IP_\rmap f = \La_\rmap f$, 
then $\La_\rmap$ is an isolated eigenvalue. 

\begin{example}\label{ex:6}
$\map_1(x) := 1/2 + {\rm sign}(x-1/2) \cdot
               [1/2 - |2{\rm sign}(x-1/2)\cdot(x-1/2) - 1/2|]$,
where ${\rm sign}(x)$ is the sign of the number $x$, and
$\map_2(x):=x/2$.
\end{example}

The maps $\map_i$ are shown on Fig.~\ref{fig:ex-is-eigen}.
Note that this random map is contractive on average when $0<p<1/3$.%
\Bfig(150,150)
      {\footnotesize{
       \bline(0,0)(1,0)(150)   \bline(0,0)(0,1)(150)
       \bline(0,150)(1,0)(150) \bline(150,0)(0,1)(150)
       \bezier{100}(0,0)(75,75)(150,150) 
       \bline(0,75)(1,-2)(37) \bline(37,0)(1,+2)(75) \bline(150,75)(-1,+2)(37)
       \put(25,25){\circle*{3}} \bezier{15}(25,25)(25,12)(25,0)
       \put(125,125){\circle*{3}} \bezier{30}(125,125)(125,75)(125,0)
       \bline(0,35)(2,1)(150) \bezier{25}(73,73)(73,36)(73,0)
       \put(23,-8){$c$} \put(123,-8){$1-c$} \put(70,-8){$1/2$}
       \put(90,130){$\map_{1}$} \put(105,80){$\map_{2}$}
      }}
{Example of a random map with a nontrivial isolated eigenvalue
\label{fig:ex-is-eigen}}
By the construction the map $\map_1$ has three unstable fixed
points $c=1/6$, $1/2$ and $1-c=5/6$ and the point $1/2$ is the only 
fixed point of the contractive map $\map_2$. Besides, the trajectories 
of the random map starting at points $c$ and $1-c$ are symmetric and 
consist of a countable number of points 
$\{2^{-1} - 2^{-n}c\}_n$ and $\{2^{-1} + 2^{-n}(1-c)\}_n$ respectively.
Consider a family of generalized functions 
$$ f_{\bar a}(x) := \sum_{k=0}^n a_k \cdot \1{2^{-1} - 2^{-k}c}(x)
                  - \sum_{k=0}^n a_k \cdot \1{2^{-1} + 2^{-k}(1-c)}(x) ,$$
parametrized by the sequence of coefficients $\bar a=\{a_k\}_k$. Due to 
the previous remark this family is invariant with respect to the 
action of $\IP_\rmap$ and our aim is to find such a sequence $\bar a$ 
that $\IP_\rmap f_{\bar a} = \La_{\rmap}f_{\bar a}$. 

\begin{lemma}\label{l:ex-is-eigen} In the example~\ref{ex:6}
for each $0<p<1/5$ there is a sequence $\bar a=\bar a(p)$ such 
that $\IP_\rmap f_{\bar a} = \La_{\rmap}f_{\bar a}$.
However $\na{f_{\bar a}}{\alpha}=\infty$ for any 
$\alpha\in(\frac12\log_2(1/p-1),1)$.
\end{lemma}

\proof Observe that $\La_{\rmap}=(1+3p)/2<1$. Therefore rewriting 
the eigenvalue relation in terms of the weights $a_k$ we obtain
the following recurrent relations:%
\bea{((1+3p)/2)\cdot a_0 \a= p a_{0} + p a_{1} \\
     ((1+3p)/2)\cdot a_k \a= (1-p) a_{k-1} + p a_{k+1} }%
for each $k\ge1$. Solving these equations with respect to the
variables with higher indices we get:
$$ a_{1}   = \frac{1+p}{2p} a_{0} , \qquad
   a_{k+1} = \frac{1+3p}{2p}a_k - \frac{1-p}p a_{k-1} 
   \qquad {\rm for}\quad k\ge1 .$$
The eigenvalues of the matrix %
$A=\pmatrix{\frac{1+3p}{2p} & -\frac{1-p}{p} \cr 1 & 0 \cr}$,
controlling the growth of the coefficients, are equal to $2$ and 
$(1-p)/(2p)$ respectively. Now since for $0<p<1/5$ the second 
eigenvalue is larger and since $a_{1}/a_{0}=(1+p)/(2p)>2$ we deduce 
that the constants $a_{k}$ grow as $((1-p)/(2p))^{k}$ as $k\to\infty$. 

It remains to estimate the $\na{\cdot}{\alpha}$-norm of the 
generalized function $f_{\bar a}$:
$$ \na{f_{\bar a}}{\alpha} 
 = 2c^\alpha \sum_{k=0}^\infty a_k 2^{-k\alpha} < \infty $$
if and only if $((1-p)/(2p))2^{-k\alpha} < 1$. On the other hand, 
$(1-p)/(2p)>2$ for $0<p<1/5$ which contradicts to the convergence. \qed

\subsection{Spectrum for the case of expanding on average random maps}

Let $\map_i$ for each $i$ be a map from the unit interval into itself
and let the following condition holds:
$$ \gamma_\rmap := \sup_x \sum_i \frac{p_i}{|\map'_i(x)|} < 1 ,$$
where the supremum is taken over all points $x\in X$ where the
derivatives of maps $\map_i$ are well defined. Then, according to
Theorem~\ref{rand-map-bounded}, the Lasota-Yorke inequality is valid:
$$ \var(\IP_{\rmap}^{n}h)
                   \le C\gamma_\rmap^{n}\var(h) + \beta||h|| .$$
Therefore all known results on the spectral properties of the
Perron-Frobenius operator obtained for piecewise expanding maps
based on a similar inequalities remain valid as well (see for example
the detailed discussion in \cite{Bl-mon}).

As we already mentioned the condition~(\ref{Mor-ineq}) does not imply
the inequality of Lasota-Yorke type and moreover there are
examples when under the condition~(\ref{Mor-ineq}) there is no
exponential correlation decay. Therefore the question how to
extend the description of the spectrum to this case remains open.

\subsection{Stochastic stability}

In this section we shall study random perturbations of random
maps under consideration. Since under the condition~(\ref{Pel-ineq})
perturbations of expanding on average random maps can be considered
exactly as in the case of deterministic piecewise expanding maps
(see \cite{BK95,BK97} and general discussion in \cite{Bl-mon})
we shall restrict ourselves only to the case of contracting on
average systems.

As usual under the randomly perturbed system we shall mean the
superposition of the original system and a Markov process acting
on the same phase space and defined by the family of transition
operators $Q_\ep$ (here $\ep$ stands for the `size' of
perturbation).

To simplify the calculations we shall start from the case
$X=\Tor^d$ and then shall explain how the corresponding arguments
should be changed in the case of a general smooth manifold.

Consider two families of operators: integral operators
$Q_\ep:\cF_\alpha\to\IC^1$ and the dual ones
$Q_\ep^*:\IC^1\to\IC^1$:
$$ Q_\ep f(x) := \int q_\ep(z,x) f(z) \, dz , \CR
   Q_\ep^* \phi(x) := \int q_\ep(x,z) \phi(z) \, dz $$
with the family of nonnegative kernels $q_\ep(\cdot,\cdot)$, with
respect to which we shall assume that for some $1<M<\infty$ the
following conditions hold:%
\beq{kernel-1}{ \int q_\ep(x,y)\, dy =1, \quad q_\ep(x,y)=0
                \quad \forall \rho(x,y)>\ep, }
\beq{kernel-2}{ \int |q_\ep(x,z) - q_\ep(y,z+y-x)|~dz \le M \rho(x,y), }

We start the analysis of the operator $Q_\ep^*$ from the following simple
estimates.

\begin{lemma}\label{l:simple-est} For each $\phi\in\IC^\alpha$ we have
\beq{est-mod}{|Q_\ep^* \phi|_\infty \le |\phi|_\infty ,}
\beq{est-dif}{|Q_\ep^* \phi - \phi|_\infty \le \ep^\alpha H_\alpha(\phi) .}
\end{lemma}

\proof The proof is straightforward:%
\bea{|Q_\ep^* \phi(x)| \a= |\int q_\ep(x,z)\phi(z)\,dz|
 \le \int q_\ep(x,z) |\phi(z)|\,dz \le |\phi|_\infty .\\
 |Q_{\ep}^{*}\phi(x) - \phi(x)|
 \a= |\int q_{\ep}(x,z)\phi(z)\,dz - \phi(x)| \\
 \a\le \int q_{\ep}(x,z) |\phi(z) - \phi(x)| \,dz \\
 \a\le \ep^{\alpha} H_{\alpha}(\phi) ,}%
since only the points $z\in B_\ep(x)$ should be taken into account.
\qed

Now let us estimate the norm of the operator $Q_\ep$.

\begin{lemma}\label{l:Qep-alpha} Let 
$M_1(\ep):=\max\{2 \ep^{\alpha/2}, \, M\ep^{(1-\alpha)/2}\}$. Then
$\na{Q_\ep}{\alpha} \le 1 + M_1(\ep)$. \end{lemma}

\proof Our aim is to show that $Q_\ep^* \phi$ is a valid test
function and to estimate the values of $V_\alpha(Q_\ep^* \phi)$.
We already estimated the supremum norm of the function $Q_\ep^*
\phi$. To get the estimate of the H\"older constant we consider
two different situations: when the points $x,y$ are close, i.e.  
$\rho(x,y)\le\sqrt{\ep}$, and the opposite case when they are far 
apart. In the first case we proceed as follows:%
\bea{|Q_\ep^* \phi(x) - Q_\ep^* \phi(y)|
 \a= |\int q_\ep(x,z) \phi(z)\,dz - \int q_\ep(y,z) \phi(z)\,dz | \\
 \a= |\int q_\ep(x,z) \phi(z)\,dz - \int q_\ep(y,z+y-x) \phi(z+y-x)\,dz | \\
 \a\le |\int q_\ep(x,z) \phi(z)\,dz - \int q_\ep(x,z) \phi(z+y-x)\,dz | \\
 \a~+ |\int q_\ep(x,z) \phi(z+y-x)\,dz - \int q_\ep(y,z+y-x) \phi(z+y-x)\,dz| \\
 \a\le \int q_\ep(x,z) |\phi(z) - \phi(z+y-x)|\,dz \\
   \a~+ \int |q_\ep(x,z) - q_\ep(y,z+y-x)|\cdot |\phi(z+y-x)|\,dz \\
 \a\le \rho^\alpha(x,y) H_\alpha(\phi) + M \rho(x,y)|\phi|_\infty  \\
 \a\le \left[H_\alpha(\phi)
   + M \ep^{(1-\alpha)/2}|\phi|_\infty \right] \rho^\alpha(x,y) .}%
In the opposite case, when $\rho(x,y)>\sqrt{\ep}$ we shall proceed in
a different way:%
\bea{|Q_\ep^* \phi(x) - Q_\ep^* \phi(y)|
 \a\le |Q_\ep^* \phi(x) - \phi(x)| + |Q_\ep^* \phi(y) - \phi(y)|
   + |\phi(x) - \phi(y)| \\
 \a\le 2|Q_\ep^* \phi - \phi|_\infty + \rho^\alpha(x,y) H_\alpha(\phi) \\
 \a\le 2\ep^\alpha H_\alpha(\phi) + \rho^\alpha(x,y) H_\alpha(\phi) \\
 \a\le (1 + 2\ep^{\alpha/2}) \rho^\alpha(x,y) H_\alpha(\phi) .}%
Hence we have%
\bea{V_\alpha(Q_\ep^* \phi) 
 \a= H_\alpha(Q_\ep^* \phi) + |Q_\ep^* \phi|_\infty \\
 \a\le (1 + 2 \ep^{\alpha/2})H_\alpha(\phi) 
   + M \ep^{(1-\alpha)/2}|\phi|_\infty \\
 \a\le (1 + \max\{2 \ep^{\alpha/2}, \, M\ep^{(1-\alpha)/2}\})V_\alpha(\phi) .}%
Thus, setting $M_1(\ep):=\max\{2 \ep^{\alpha/2}, \, M\ep^{(1-\alpha)/2}\}$, 
we get%
\bea{\na{Q_\ep f}{\alpha}
 \a= \sup_{V_{\alpha}(\phi)\le} \int Q_\ep f \cdot \phi
   = \sup_{V_{\alpha}(\phi)\le} \int f \cdot Q_\ep^* \phi \\
 \a\le \sup_{V_{\alpha}(\phi)\le1} V_{\alpha}(Q_\ep^* \phi) \cdot
     \sup_{V_{\alpha}(\psi)\le1} \int f \cdot \psi \\
 \a\le (1 + M_1(\ep)) \cdot \na{f}{\alpha} .}%
Observe that in several places we used the estimates of the supremum norms
obtained in Lemma~\ref{l:simple-est}.
\qed

\begin{lemma}\label{l:3norm-0} Let $G:\cF_{\alpha}\to\cF_{\alpha}$
be a linear operator, and let $G^{*}:\IC^{1}\to\IC^{1}$ be dual
to it, i.e. $\int Gf\cdot\phi = \int f\cdot G^{*}\phi$. Then for
all $f\in\cF_{\beta}$
$$ \na{Gf}{\beta}
   \le \left(\sup_{V_{\beta}(\phi)\le1} V_{\alpha}(G^{*}\phi) \right)
       \cdot \na{f}{\alpha} .$$
\end{lemma}

\proof Indeed,
\bea{\na{Gf}{\beta}
 \a= \sup_{V_{\beta}(\phi)\le1} \int Gf\cdot\phi
 = \sup_{V_{\beta}(\phi)\le1} \int f\cdot G^{*}\phi  \\
 \a\le \left(\sup_{V_{\beta}(\phi)\le1} V_{\alpha}(G^{*}\phi) \right)
       \cdot \sup_{V_{\alpha}(\psi)\le1} \int f\cdot\psi .}%
\qed

\begin{lemma}\label{l:3norm}
$$ |||Q_{\ep} - 1||| \equiv \na{Q_{\ep} - 1}{\beta\to\alpha} :=
   \sup_{\na{f}{\alpha}\le1}\na{Q_{\ep}f - f}{\beta} \to 0
       \quad {\rm as} \quad \ep\to0.$$
\end{lemma}

\proof Applying Lemma~\ref{l:3norm-0} to the operator
$G=Q_{\ep}-1$ we get that the sufficient condition of the
validity of the desired statement is the convergence of
$$ \sup_{V_{\beta}(\phi)\le1} V_{\alpha}(Q_{\ep}^{*}\phi - \phi)
   \to 0 $$
as $\ep\to0$. Let us prove this convergence. Observe that since
$\beta>\alpha$ and $\phi\in\IC^\beta$ we can get a
stronger estimate compare to Lemma~\ref{l:simple-est}:%
\bea{|Q_{\ep}^{*}\phi(x) - \phi(x)|
 \a= |\int q_{\ep}(x,z)\phi(z)\,dz - \phi(x)| \\
 \a\le \int q_{\ep}(x,z) |\phi(z) - \phi(x)| \,dz \\
 \a\le \ep^{\beta} H_{\beta}(\phi) .}%
Applying now estimates similar to ones used in the proof of
Lemma~\ref{l:Qep-alpha} and taking into account that we consider
more smooth test-functions $\phi\in\IC^{\beta}$ in the case 
$\rho(x,y)\le\ep$ we get%
\bea{\a\left|\left(Q_{\ep}^{*}\phi(x) - \phi(x)\right)
       - \left(Q_{\ep}^{*}\phi(y) - \phi(y)\right) \right| \\
 \a~\le |Q_{\ep}^{*}\phi(x) - Q_{\ep}^{*}\phi(y)| + |\phi(x) - \phi(y)| \\
 \a~\le \rho^{\beta}(x,y) H_{\beta}(\phi) + M\rho(x,y)|\phi|_{\infty}
   + \rho^{\beta}(x,y) H_{\beta}(\phi) \\
 \a~\le \left[2 \ep^{\beta-\alpha}H_{\beta}(\phi)
     + M\ep^{1-\alpha}|\phi|_\infty \right] \rho^\alpha(x,y) .}%
While in the opposite case, when $\rho(x,y)>\ep$, using the same
argument as in the proof of Lemma~\ref{l:Qep-alpha} we get%
\bea{|\left(Q_{\ep}^{*}\phi(x) - \phi(x)\right)
       - \left(Q_{\ep}^{*}\phi(y) - \phi(y)\right)|
 \a\le 2 |Q_{\ep}^{*}\phi - \phi|_\infty
   \le 2 \ep^{\beta} H_{\beta}(\phi) \\
 \a\le 2 \ep^{\beta-\alpha}H_{\beta}(\phi) \rho^\alpha(x,y) .}%
Hence for $\phi\in\IC^{\beta}$ 
$$ H_{\alpha}(Q_{\ep}^{*}\phi - \phi)
 \le 2 \ep^{\beta-\alpha}H_{\beta}(\phi)
   + M \ep^{1-\alpha} |\phi|_{\infty} ,$$
which yields the following estimate%
\bea{V_{\alpha}(Q_{\ep}^{*}\phi - \phi)
 \a\le 2 \ep^{\beta-\alpha}H_{\beta}(\phi)
   + M \ep^{1-\alpha} |\phi|_{\infty} + \ep^{\beta} H_{\beta}(\phi) \\
 \a\le 3 \ep^{\beta-\alpha}H_{\beta}(\phi)
   + M \ep^{1-\alpha} |\phi|_{\infty}
 \le (3 + M\ep^{1-\beta})\ep^{\beta-\alpha} V_{\alpha}(\phi) \to 0 }%
as $\ep\to0$. \qed

The properties of the transition operator obtained above together
with Theorem~\ref{th-LY-rand-contr} under the additional
assumption that $\La_\rmap(\alpha)<1/2$ make it possible to use
results about the spectral stability of transfer operators
satisfying Lasota-Yorke type inequalities \cite{KeLi} and to
obtain the following stability result.

\begin{theorem}\label{t:stoch-stab} Let the conditions
(\ref{kernel-1}, \ref{kernel-2}) be satisfied and let
$\La_\rmap(\alpha)<1/2$ for some $\alpha\in(0,1)$. Then all
elements of the spectrum $\sp_{\cF_{\alpha}}(\IP_{\rmap})$
outside of the disk of radius $\La_\rmap(\alpha)$ are
stochastically stable and the corresponding eigenprojectors of
the perturbed system converge to the genuine ones.
\end{theorem}

\proof First let us show that the transfer operator for the
stochastically perturbed system satisfies a Lasota-Yorke type
inequality. A straightforward calculation shows that this
operator is equal to $Q_\ep \IP_\rmap$. Combining the results of
Lemma~\ref{l:Qep-alpha} and Theorem~\ref{th-LY-rand-contr} we
get for any $h\in\cF_\alpha$ that%
\bea{\na{Q_\ep \IP_\rmap h}{\alpha}
   \a\le (1 + M_1(\ep))\na{\IP_\rmap h}{\alpha} \\
   \a\le (1 + M_1(\ep)) \kappa \La_\rmap(\alpha) \na{h}{\alpha}
     + \const\cdot(\kappa-2)^{-1/\alpha}\na{h}{\beta} .}%
Therefore, if $\La_\rmap(\alpha)<2$ for some $0<\alpha<1$, then the
number $\gamma:=(1 + M_1(\ep)) \kappa \La_\rmap(\alpha) <1$ for
the value of $\kappa>2$ guaranteed by Theorem~\ref{th-LY-rand-contr}.
Since all other assumptions of the abstract spectral stability
result in \cite{KeLi} were alredy checked during the analysis of
our Banach spaces of generalized functions, we come to the desired
statement. \qed

To proceed further we need to generalize the notion of the periodic
turning point, well known in the one-dimensional dynamics. Namely,
a point $x\in X$ is called the {\em periodic turning point} for the
map $\map:X\to X$ if $\map^nx=x$ for some $n\in\IZ_+$ and the
derivative of the map $\map$ is not well defined at the point $x$.

\begin{definition} A point $x\in X$ is called the {\em periodic turning
point} for the {\em random} map $\rmap:X\to X$ if there is a finite
collection of indices $i_1,i_2,\dots,i_k$ such that the point $x$ is
the periodic turning point for the deterministic map
$\map_{i_1}\circ\map_{i_2}\circ\dots\map_{i_k}$.
\end{definition}

For example, the point $x=1/2$ is the periodic turning point for
the random map in the example 2 for any nontrivial distribution ($0<p<1$).

\begin{theorem}\label{c:drop} The assumption $\La_\rmap(\alpha)<1/2$
can be replaced by the following: either all the maps $\map_{i}$
are bijective and $C^1$-differentiable, or they are piecewise
$C^1$-differentiable and have no periodic turning points. Then all
isolated eigenvalues are stochastically stable.
\end{theorem}

\proof The key idea here is to consider another representation of
the perturbed operator %
$(Q_{\ep}\IP_{\rmap})^{n} = \tilde Q_{\ep(n)}\IP_{\rmap}^{n}$ %
and to show that the new operator $\tilde Q_{\ep(n)}$ satisfies
the same assumptions (\ref{kernel-1}, \ref{kernel-2}) as the
operator $Q_{\ep}$ (except that the value of $\ep(n)$ might be in
$1/\min\{\La_{\map_i}\}$ times larger). Therefore choosing $n$
large enough we can always get $\La_\rmap^{n}(\alpha)<1/2$. This
idea was first applied in the case of piecewise expanding maps in
\cite{BK95,BK97} and then in the case of hyperbolic maps in
\cite{BKL}. If all maps $\map_{i}$ are bijective this can be done
by a simple change of variables, while in the second case the
construction is more involved but is completely similar to the
one in \cite{BK95,BK97}. \qed

Let us show now what should be changed in the case of a general
smooth manifold. Since locally in a neighborhood of a point $x\in
X$ one can introduce local coordinates by means of the
exponential map $\Psi_x$, the tangent linear space ${\cal T}_x X$
can be isometrically mapped into $\IR^d$. Let $\nu>0$ be a number
such that for each point $x\in X$ the ball (v the metrics $\rho$)
of radius $\nu$ centered at this point belongs to the domain of
values of the exponential map $\Psi_x$. Note that we already have
introduced the restriction on the distance between the points in
the definition of the H\"older constant needing to be content
with the domain of definition of the exponential map. In fact,
the first difference appears only in the analysis of random
perturbations, in particular, the condition (\ref{kernel-2})
should be rewritten as%
$$ \int|q_\ep(x,y) - q_\ep(\Psi_x(\Psi_x^{-1}(x)+t),\Psi_y(\Psi_y^{-1}(y)+t))|
 ~dy \le \rho(x, \Psi_x(\Psi_x^{-1}(x)+t))M ,$$
where $t\in\IR^d$ and $|t|\le\nu$.

Assuming now that $\ep<\nu$ and replacing the expressions of type
$z+y-x$ to
$$ \Psi_z\left(\Psi_z^{-1}(z) + \Psi_z^{-1}(y) - \Psi_z^{-1}(x)\right) ,$$
we obtain the same estimates as in the flat case (when $X$ is the
unit torus). Therefore all results of this section remain valid
for the case of a general smooth manifold.

\subsection{Finite rank approximations}

Let us discuss now finite dimensional approximations of transfer
operators. Again due to the same reason as in the previous
section we shall restrict the analysis to the case of contracting
on average random maps.

Let $\{\Delta_{i}\}_{i}$ be a finite partitions of the phase
space $X$ into domains (cells) $\Delta_{i}$ of diameter not
larger than $\delta>0$. For a point $x\in X$ by $\Delta_{x}$ we
denote the element of the partition containing it. Under these
notation the so called Ulam approximation can be described as an
operator
$$ \tilde Q_{\delta}f(x) := \frac1{|\Delta|}\int_{\Delta_{x}}f .$$
Note that this operator is selfdual, i.e. $\tilde Q_{\delta} =
\tilde Q_{\delta}^*$. One can easily check also that the
dimension of the space $\tilde Q_{\delta}\cF_{\alpha}$ coincides
with the number of elements in the Ulam partition.

\begin{lemma} $\na{Q_{\delta}}{\alpha}=\infty$. \end{lemma}

\proof Let a point $y_0\in X$ belongs to the boundary between two
elements of the Ulam partition, and let the points $y(\ep)$ and
$y'(\ep)$ belong to neighboring elements of the partition both on
the distance $\ep$ from $y_0$. For the function
$$ f_\ep(x):=\1{y(\ep)}(x) + \1{y'(\ep)}(x) ,$$
where $\1{y}$ means the $\delta$-function at the point $y$, the
following inequalities hold:
$$ \na{f_\ep}{\alpha} \le \const\ep^\alpha ,\CR
   \na{Q_{\delta}f_\ep}{\alpha} \ge \const > 0 .$$
The first of these inequalities follows from the definition of
the norm $\na{\cdot}{\alpha}$, while the second one is a
consequence of the fact that the function $Q_{\delta}f_\ep$ is
the characteristic function of the union of two neighboring
elements of the partition containing the points $y(\ep)$ and 
$y'(\ep)$. Thus,
$\na{Q_{\delta}f_\ep}{\alpha}/\na{f_\ep}{\alpha} \to \infty$
as $\ep\to0$. \qed

This result shows that the original Ulam approximation scheme
cannot be immediately applied to the spectral analysis of random
maps.

What is still possible is that the leading eigenfunction -- \SBR
measure may be stable for the class of maps we consider (the
above example does not contradict to this -- the \SBR measure is
preserved). I believe there should be very deep reasons
explaining the stability of the leading eigenfunction, while all
others are not stable, however presently we do not have the
adequate explanation. In the literature (see, for example,
\cite{Bl-mon} and further references therein) the stability of the
\SBR measure is proven for the class of piecewise expanding maps.
Moreover numerous numerical studies confirm this stability for a
much broader class of dynamical systems. To the best of our
knowledge the following simple example of a one-dimensional
discontinuous map represent the first counterexample to the
original Ulam hypothesis.

\begin{lemma}\label{l:counter-or-ulam} The map
$$ \map x := \function{
            \frac{x}4+\frac12 &\mbox{if }\; 0\le x <\frac5{12}\\
            -2x +1            &\mbox{if }\; \frac5{12}\le x < \frac12\\
            \frac{x}2+\frac14 &\mbox{otherwise} .} $$
from the unit interval into itself is uniquely ergodic, but the
leading eigenvector of the Ulam approximation
$\Pi_{1/n}\IP_{\map}$ does not converge weakly to the only
$\map$-invariant measure.
\end{lemma}

\Bfig(150,150)
      {\footnotesize{
       \bline(0,0)(1,0)(150)   \bline(0,0)(0,1)(150)
       \bline(0,150)(1,0)(150) \bline(150,0)(0,1)(150)
       \bdline[6,8](0,75)(1,0)(150)  \bdline[6,8](75,0)(0,1)(150)
       \put(0,75){\vector(4,1){60}}  \put(60,30){\vector(1,-2){15}}
       \bline(75,75)(2,1)(75)
       \bdline[6,8](60,90)(0,-1)(90)
       \put(-3,-8){$0$} \put(147,-8){$1$} \put(50,-8){$5/12$} \put(72,-8){$1/2$}
      }}
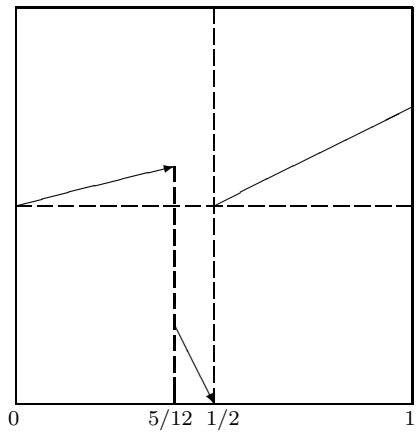
{Counterexample for the original Ulam construction.
\label{fig-counter-ulam}}

Observe that the situation when the \SBR measure becomes unstable
with respect to Ulam scheme is indeed very exotic and the map in
the corresponding example is not only discontinuous (see
Fig.~\ref{fig-counter-ulam}), but this discontinuity occurs in a
periodic turning point (compare to instability results about
general random perturbations in \cite{BK97}).

\proof Denote by $v^{(n)}$ the normalized leading eigenvector of
the matrix $\Pi_{1/n}\IP_{\map}$. A straightforward calculation
shows that for each $n\in\IZ_{+}^{1}$ all entries of the vector
$v^{(2n+1)}$ are zeros except the first entry, which is equal to
$1/3$ and the $(n+1)$-th one, which is equal to $2/3$. Compare
this to the only invariant measure of the map $\map$ -- the unit
mass at the point $1/2$. \qed

To overcome this difficulty we consider another `smoothed'
approximation scheme. In each element of the partition $\{\Delta_i\}$
(of diametr $\le\delta$) we fix an arbitrary point (its `center')
$x_i\in\Delta_{i}$. Now for a given smooth enough kernel
$q_\ep(\cdot,\cdot)$ satisfying the assumptions from the previous
section we define the following finite dimensional operator:
$$ Q_{\ep,\delta} f(x)
 := \sum_i 1_{\Delta_i}(x) \int q_\ep(z,x_i) f(z)~dz .$$
Observe that the dual operator is equal to:
$$ Q_{\ep,\delta}^{*} \phi(x)
 := \sum_i q_\ep(x, x_i) \int_{\Delta_i} \phi(z)~dz .$$
Indeed,%
\bea{\int Q_{\ep,\delta} f(x) \cdot \phi(x)~dx
 \a= \int \sum_i 1_{\Delta_i}(x) \int q_\ep(z,x_i) f(z)~dz \cdot \phi(x)~dx \\
 \a= \int f(z) \sum_i q_\ep(z,x_i)
               \left(\int 1_{\Delta_i}(x) \phi(x)~dx \right)~dz \\
 \a= \int f(z) \sum_i q_\ep(z,x_i) \int_{\Delta_i} \phi(x)~dx~dz
   = \int f(z) \cdot Q_{\ep,\delta}^{*} \phi(z)~dz .}%

\begin{lemma}\label{l:smooth-ulam-s} Let additionally to the assumptions
(\ref{kernel-1}, \ref{kernel-2}) for any points $x,y,z\in X$
the inequality%
\beq{kernel-4}{ |q_\ep(x,y) - q_\ep(x,z)| + |q_\ep(y,x) - q_\ep(z,x)|
      \le M \ep^{-d-1} \rho(y,z) . }
holds. Then
$$ \na{Q_\ep f - Q_{\ep,\delta} f}{\alpha}
   \le 5M \ep^{-d-1} \delta^{1-\alpha} \na{f}{\alpha} .$$
\end{lemma}

\proof
$$ \int (Q_\ep f - Q_{\ep,\delta} f)\cdot \phi
 = \int f \cdot (Q_\ep^* \phi - Q_{\ep,\delta}^* \phi) .$$
Denote
$$ \Phi(x)
 := Q_\ep^* \phi(x) - Q_{\ep,\delta}^* \phi(x)
 = \sum_i \int_{\Delta_i} (q_\ep(x,z) - q_\ep(x,x_i)) \phi(z)~dz .$$
Then%
\bea{|\Phi|_\infty
 \a\le |\phi|_\infty \cdot
     \sup_x \sum_i \int_{\Delta_i} |q_\ep(x,z) - q_\ep(x_i,x)|~dz \\
 \a\le |\phi|_\infty \cdot
     \sup_x \sum_i \int_{\Delta_i}
     \left(|q_\ep(x,z) - q_\ep(x,x_i)|
         + |q_\ep(x,x_i) - q_\ep(x_i,x)| \right)~dz \\
 \a\le 2M \ep^{-d-1} \delta |\phi|_\infty ,}%
since $x,z\in\Delta_i$ and, hence,
$\max\{\rho(z,x_i),\rho(x,x_i)\}\le\delta$.

Let us estimate $H_\alpha(\Phi)$. If $\rho(x,y) \le \delta$ then%
\bea{|\Phi(x) - \Phi(y)|
 \a\le |Q_\ep^* \phi(x) - Q_\ep^* \phi(y)|
   + |Q_{\ep,\delta}^* \phi(x) - Q_{\ep,\delta}^* \phi(y)| \\
 \a\le \int|q_{\ep}(x,z) - q_{\ep}(y,z)|\cdot|\phi(z)|~dz
   + \sum_i |q_\ep(x,x_i) - q_\ep(y,x_i)| \int_{\Delta_i}|\phi| \\
 \a\le M\ep^{-d-1}\rho(x,y)|\phi|_{\infty}
   + M\ep^{-d-1} \rho(x,y)\cdot|\phi|_\infty
 = 2M\ep^{-d-1} \delta^{1-\alpha} \rho^\alpha(x,y)\cdot|\phi|_\infty .}%
Otherwise if $\rho(x,y) > \delta$ we apply another estimate
$$ |\Phi(x) - \Phi(y)| \le 2|\Phi|_\infty
 \le 4M \ep^{-d-1} \delta |\phi|_\infty
 \le 4M \ep^{-d-1} \rho^\alpha(x,y)\cdot \delta^{1-\alpha}
                       \cdot|\phi|_\infty .$$
Thus,
$$ H_\alpha(\Phi) \le 4M \ep^{-d-1} \delta^{1-\alpha}
                         \cdot|\phi|_\infty ,$$
and hence
$$ V_\alpha(\Phi) \le 5M \ep^{-d-1} \delta^{1-\alpha}
                         \cdot V_\alpha(\phi) ,$$
which yields the desired statement. \qed

\begin{theorem}\label{smooth-ulam-s} Let the family of kernels
$\{q_\ep(\cdot,\cdot)\}$ satisfies the conditions
(\ref{kernel-1}, \ref{kernel-2},\ref{kernel-4}). Then%
\bea{\na{Q_{\ep,\delta}}{\alpha}
                    \a\le 1 + M_1(\ep) + 5M \ep^{-d-1} \delta^{1-\alpha}, \\
     |||Q_{\ep,\delta} - 1||| 
                    \a\le (5M + 3 + M\ep^{1-\beta})
                          (\ep^{-d-1} \delta^{1-\alpha} 
                        + \ep^{\beta-\alpha}) \to 0
                    \quad {\rm as} \;
                    \ep^{-d-1} \delta^{1-\alpha} + \ep^{\beta-\alpha}\to0 .}%
Hence for the case $\La_\rmap(\alpha)<1/2$ the isolated
eigenvalues and the corresponding eigenprojectors of the operator
$\IP_\rmap$ are stable with respect to the considered
approximation.
\end{theorem}

\proof According to Lemmas \ref{l:Qep-alpha} and \ref{l:smooth-ulam-s}
$$ \na{Q_{\ep,\delta}}{\alpha}
 \le \na{Q_{\ep}}{\alpha} + \na{Q_{\ep,\delta} - Q_{\ep}}{\alpha}
 \le 1 + M_1(\ep) + 5M \ep^{-d-1} \delta^{1-\alpha} ,$$
which proves the first statement.

Similarly but using Lemma~\ref{l:3norm} instead of
Lemma~\ref{l:Qep-alpha}, we get%
\bea{V_\alpha(Q_{\ep,\delta}\phi - \phi)
 \a\le V_\alpha(Q_{\ep,\delta}\phi - Q_{\ep}\phi)
   + V_\alpha(Q_{\ep}\phi - \phi) \\
 \a\le \left(5M \ep^{-d-1} \delta^{1-\alpha} 
   + (3 + M\ep^{1-\beta})\ep^{\beta-\alpha}\right) 
       \cdot V_\alpha(\phi) ,}%
which finishes the proof. \qed

In fact the finite dimensional approximation defined by the
two-parameter family of operators
$\{Q_{\ep,\delta}\}_{\ep,\delta}$ one can consider as a smoothed
version of the original Ulam construction, which corresponds to
the case $\ep=0$. Observe that in our approximations the relation
between the parameters is completely different -- it is necessary
that $\ep \gg \delta$.

We consider also another (seeming more natural) finite rank
approximation scheme. Denote by $\Pi_\delta$ the pure Ulam
approximation operator corresponding to the partition into
domains $\{\Delta_i\}$ whose diameters do not exceed $\delta$:
$$ \Pi_\delta f(x) := \frac1{|\Delta_x|} \int_{\Delta_x} f(s)~ds ,$$
where $\Delta_x$ stands for the element of the partition
containing the point $x$. Note that this operator is self
adjoint. We shall approximate our transfer operator $\IP_\rmap$
by $\Pi_\delta Q_\ep \IP_\rmap$. To study the properties of this
approximation we need as usual to analyze properties of the
adjoint operator, i.e. of the operator
$$ Q_\ep^* \Pi_\delta^* \phi(x)
 = \int q_\ep(x,z) \frac1{|\Delta_z|} \int_{\Delta_x} \phi(s)~ds~dz .$$

\begin{lemma}\label{l:f-r-comp}
$\na{Q_\ep - \Pi_\delta Q_\ep}{\alpha}
   \le (3+2M) \ep^{-d-1} \delta^{\alpha(1-\alpha)} \to 0$ 
as $\ep^{-d-1} \delta^{\alpha(1-\alpha)}\to0$.
\end{lemma}

\proof Denote%
\bea{\Phi(x) \a:= Q_\ep^*\phi(x) - Q_\ep^* \Pi_\delta^* \phi(x)
 = \int q_\ep(x,z)
  \left( \phi(z) - \frac1{|\Delta_z|}\int_{\Delta_x} \phi(s)~ds\right) ~dz \\
 \a= \int q_\ep(x,z) \frac1{|\Delta_z|}
     \int_{\Delta_x} (\phi(z) - \phi(s))~ds~dz .}%
Since $\phi\in\IC^\alpha$ and the diameter of the elements of the
partition does not exceed $\delta$, we have
$$ |\Phi|_\infty \le \delta^\alpha H_\alpha(\phi) \sup_x \int q_\ep(x,z)~dz
 = \delta^\alpha H_\alpha(\phi) .$$
Now we are going to estimate the H\"older constant of the function $\Phi$.
which we shall do in two steps. First, we consider the case when
$\rho(x,y)\le\delta^\alpha$:%
\bea{|\Phi(x) - \Phi(y)|
 \a\le |Q_\ep^*\phi(x) - Q_\ep^*\phi(y)|
     + |Q_\ep^* \Pi_\delta^* \phi(x) - Q_\ep^* \Pi_\delta^* \phi(y)| \\
 \a\le \int |q_\ep(x,z) - q_\ep(y,z)|\cdot|\phi|_\infty~dz
     + \int |q_\ep(x,z) - q_\ep(y,z)|\cdot
       \frac1{|\Delta_z|}\int_{\Delta_z}|\phi(s)|~ds~dz \\
 \a\le 2M\ep^{-d-1}\rho(x,y) |\phi|_\infty
   \le 2M\ep^{-d-1}\delta^{\alpha(1-\alpha)} |\phi|_\infty \rho^\alpha(x,y) .}%
In the opposite case, when $r(x,y)>\delta^\alpha$ we use a
different estimate:
$$ |\Phi(x) - \Phi(y)| \le 2|\Phi|_\infty
 \le 2\delta^\alpha H_\alpha(\phi)
 \le 2\delta^{\alpha(1-\alpha)}H_\alpha(\phi) \rho^\alpha(x,y) .$$
Thus %
\bea{V_\alpha(\Phi)
 \a\le 2\delta^{\alpha(1-\alpha)} H_\alpha(\phi)
     + 2M \ep^{-d-1}\delta^{\alpha(1-\alpha)} |\phi|_\infty
     + \delta^\alpha H_\alpha(\phi)  \\
 \a\le 3\delta^{\alpha(1-\alpha)} H_\alpha(\phi)
     + 2M \ep^{-d-1}\delta^{\alpha(1-\alpha)} |\phi|_\infty
   \le (3+2M) \ep^{-d-1}\delta^{\alpha(1-\alpha)} V_\alpha(\phi) \to 0 }%
as $\ep^{-d-1}\delta^{\alpha(1-\alpha)} \to 0$. \qed

Observe that the rate of convergence in this approximation is lower
compare to the previous one, however the numerical application of the 
2nd scheme is more straitforward.

\begin{corollary} Again as in Theorem~\ref{c:drop} and due to the
same reason the assumption $\La_\rmap(\alpha)<1/2$ can be
replaced by either the bijectivity of the maps $\map_{i}$ or
the absence of its periodic turning points.
\end{corollary}


\end{document}